\documentclass[%
 reprint,%
 amssymb, amsmath,%
 aip,cha,%
]{revtex4-1}
\usepackage{graphicx}
\usepackage{docs}%
\usepackage{bm}%
\usepackage[colorlinks=true,linkcolor=blue]{hyperref}%
\expandafter\ifx\csname package@font\endcsname\relax\else
 \expandafter\expandafter
 \expandafter\usepackage
 \expandafter\expandafter
 \expandafter{\csname package@font\endcsname}%
\fi
\begin{document}

\title{Fe-spin reorientation in PrFeAsO : Evidences from resistivity and specific heat studies}%
\author{D. Bhoi}
\email{dilipkumar.bhoi@saha.ac.in}
\author{P. Mandal}%
 \email{prabhat.mandal@saha.ac.in}
\affiliation{Saha Institute of Nuclear Physics, 1/AF Bidhannagar,Calcutta 700 064, India
}%
\author{P. Choudhury}

\affiliation{Central Glass and Ceramic Research Institute, 196 Raja S. C. Mullick Road, Calcutta  700 032, India
}%
\author{S. Pandya}
\author{V.Ganesan}
\affiliation{UGC-DAE Consortium for Scientific Research, University Campus, Khandwa Road, Indore 452
017, India}

\date{\today}%
\begin{abstract}
We report the magnetic field dependence of resistivity ($\rho$) and specific heat ($C$) for the non-superconducting PrFeAsO compound. Our study shows a hitherto
unobserved anomaly at $T_{SR}$ in the resistivity and specific heat data which arises as a result of the interplay of antiferromagnetic (AFM) Pr and Fe sublattices. Below the AFM transition temperature ($T_N^{\rm{Pr}}$), Pr moment orders along the crystallographic $c$ axis and its effect on the iron subsystem causes a reorientation of the ordered inplane Fe moments in a direction out of the $ab$ plane. Application of magnetic field introduces disorder in the AFM Pr sublattice, which, in turn, reduces the out-of-plane Pr-Fe exchange interaction responsible for Fe spin reorientation. Both in $\rho$($T$) and $d(C/T)/dT$ curves, the peak at $T_{SR}$ broadens with the increase of $H$ due to the introduction of the disorder in the AFM Pr sublattice by magnetic field. In $\rho$($T$) curve, the peak shifts towards lower temperature  with $H$ and disappears above 6 T while in $d(C/T)/dT$ curve the peak remains visible up to 14 T. The broadening of the anomaly at $T_N^{\rm{Pr}}$ in $C(T)$ with increasing $H$  further confirms that magnetic field induces disorder in the AFM Pr sublattice.
\end{abstract}

\maketitle
\section{Introduction}
Superconductivity in Fe-pnictides has attracted much interest because of the high critical temperature $T_c$ and the interplay of multi-band superconductivity and antiferromagnetism  mediated by the magnetic Fe ions  \cite{hosno}. On cooling, the undoped parent compounds $R$FeAsO ($R$ = La, Ce, Pr, Nd and Sm) exhibit a structural phase transition from tetragonal to orthorhombic symmetry at $T_S$ followed by an antiferromagnetic transition at $T_N$. In this family of pnictides,
the magnetic transition is well separated from the structural distortion. Neutron diffraction and several local probe techniques such as M\"{o}ssbauer spectroscopy and muon spin relaxation ($\mu$SR) show that below $T_N$ these parent compounds undergo a commensurate  AFM spin-density wave (SDW) ordering with a strongly reduced Fe moment compared to metallic iron, which does not vary significantly within the series $R$FeAsO \cite{cruz,klauss,carlo,berh,maeter}.
The Fe spins are directed along the orthorhombic $a$-axis, coupled ferromagnetically along the $b$-axis and antiferromagnetically along the $a$ axis.
A spontaneous magnetic ordering of the rare-earth  moments is also observed below $T_{N}^{R}$ = 4.4, 11, 6 and 4.66 K for $R$ = Ce \cite{zhao}, Pr \cite{kimber}, Nd \cite{tian} and Sm \cite{ding}, respectively. In CeFeAsO, the Ce moments order nearly in the $ab$ plane \cite{zhao,chen}, in PrFeAsO, the Pr moments order
along the $c$ axis \cite{zhao1}, and in NdFeAsO, the Nd moments are canted out of the $ab$ plane \cite{tian,qiu}. This static magnetic ordering of the rare-earth moments is little affected by charge doping in the  conducting Fe layer and coexists with superconductivity \cite{drew}. The coexistence of magnetic ordering and superconductivity is observed in heavy fermion systems, where the interplay between $R$ moments and conduction electrons leads to several unusual phenomena.
In oxypnictides too, the interplay between Fe and $R$ ordered moments has been reported for $R$ = Ce \cite{chi}, Nd \cite{qiu}, Pr \cite{kimber} and
Sm \cite{maeter}. In the case of PrFeAsO, the M\"{o}ssbauer data below $T_{N}^{\rm{Pr}}$ indicate the occurrence of a Fe spin reorientation phenomenon due to
the strong competition between Pr and Fe magnetic sublattices, while the Fe ordering remains unaltered below $T_{N}^{R}$ in the Sm and Ce compounds \cite{mcguire1}. So far there is no study on the effect of this Fe spin reorientation on electronic transport and thermodynamic properties. One way to visualize the influence of the Pr sublattice on the Fe sublattice is to study the electrical transport below $T_{N}^{\rm{Pr}}$, because, any reordering of the Fe moments caused by the ordering of the Pr sublattice may lead to a change in the scattering process in the Fe-layers which, in turn, influence the charge conduction phenomenon. Indeed, we observe a hitherto unknown anomaly at $T_{SR}$ in the temperature dependence of resistivity ($\rho$) and specific heat ($C$) curves of PrFeAsO below $T_{N}^{\rm{Pr}}$ and identify the same as an outcome of the interplay of the magnetic sublattices.

In this report, we present the effect of magnetic field on the abovementioned magnetic and structural phase transitions from the resistivity, Hall coefficient
and specific heat measurements. The magnetic field does not affect $T_S$ and $T_N$. However, the resistivity, specific heat and magnetic transitions at low
temperatures are significantly affected by the magnetic field. Most importantly, the maximum at $T_{SR}$ in the $\rho$ vs $T$ curve gets suppressed with increasing magnetic field and disappears above a critical value of field due to the field induced disordering in the AFM Pr-sublattice. The peak in $d(C/T)/dT$ curve at $T_{SR}$ also broadens with the increase of $H$ but it remains noticeable up to 14 T. For $R$FeAsO, such anomaly in resistivity and specific heat below the ordering temperature of rare-earth moments has not been reported so far by others. The obtained results for PrFeAsO are also compared and contrasted with those of $R$FeAsO compounds.

\section{Experimental details}

Polycrystalline PrFeAsO sample has been prepared by standard solid state reaction method using high purity chemicals Pr, As, Fe and Fe$_2$O$_3$. The details of
preparation technique is described in our earlier reports \cite{bhoi}. The phase purity of the sample was checked by powder x-ray diffraction method with
Cu$K_\alpha$ radiation and no trace of the impurity phase has been detected. The x-ray diffraction pattern can be well indexed on the basis of the tetragonal
ZrCuSiAs-type structure with the space group $P4/nmm$. The scanning electron microscope (SEM) (SUPRA, 35 VP, Carl Zeiss) image reveals well connected platelet
crystallites. The energy dispersive x-ray (EDX) analysis was used to determine the chemical composition of the sample. The EDX spectra obtained from grains of
different size and morphology reveal that the stoichiometric ratio Pr:Fe:As:O is close to the nominal composition PrFeAsO. The electrical resistivity and the
specific heat measurements were carried out in 14 T physical property measurement system (Quantum Design). The electrical resistivity was measured by keeping
the temperature fixed and sweeping the magnetic field and vice versa. Specific heat measurements were performed using relaxation technique in the temperature
range 2-200 K at different magnetic fields. The Hall coefficient ($R_H$) was determined by measuring the transverse resistivity at some selected temperatures by
sweeping the field from -7 T to 7 T in a superconducting magnet (Oxford Instruments).

\section{Results and Discussion}
\subsection{Resistivity}
The resistivity for the PrFeAsO sample is shown in Fig.1. Several magnetic and structural transitions are reflected in the $T$ dependence of $\rho$. Upon cooling, the resistivity decreases slowly down to 214 K, indicating metallic behavior. This is followed by a broad peak associated with both the structural phase transition from tetragonal to orthorhombic symmetry at $T_S$ and the spin-density wave magnetic transition at $T_N$. No thermal hysteresis has been observed in the resistivity data in the vicinity of the transitions. Because of the smooth and continuous change in resistivity around these transitions, it is difficult to
determine the individual transition temperatures from the resistivity curve. In such a case, an indicator of the transition is the temperature derivative of the
resistivity as shown in the lower inset of Fig. 1. Two overlapping peaks are identified in the d$\rho$/d$T$ vs $T$ plot, one due to the crystallographic distortion at $T_S$ (147 K) and the other due to the onset of long-range magnetic ordering of Fe moments at $T_{N}$ (140 K). Below $T_{N}$, the resistivity of the sample initially decreases with the decrease of $T$ and then passes through a shallow minimum at $T_{min} \sim$ 23 K (upper inset of Fig. 1). As $T$ is lowered further, $\rho$ exhibits a weak upturn before a sharp drop of $\sim$ 4$\%$ occurs due to the AFM ordering of 4$f$ electrons of the Pr ion. This drop in $\rho$ is appreciably larger as compared to those observed in other $R$FeAsO compounds with magnetic rare-earth ion\cite{tian,jesche,cheng,mart}. The transition
temperature $T_{N}^{\rm{Pr}}$ determined from the peak in the d$\rho$/d$T$ vs $T$ curve is $\sim$ 12 K, which is close to that reported from neutron diffraction, magnetic susceptibility and other measurements \cite{kimber,rotundu}. The increase of resistivity above the N\'{e}el temperature has also been observed in metallic $R$CuAs$_2$ for $R$ = Sm, Gd, Tb and Dy \cite{sampath}. In these compounds the increase of $\rho$ is magnetic in origin because, their nonmagnetic counterparts do not show any such behavior. However, the observation of an upturn in $\rho$ at low temperature for the nonmagnetic LaFeAsO indicates that in pnictides the origin might not be solely magnetic in nature \cite{dong}.
\begin{figure}[t]
\includegraphics[width=0.45\textwidth]{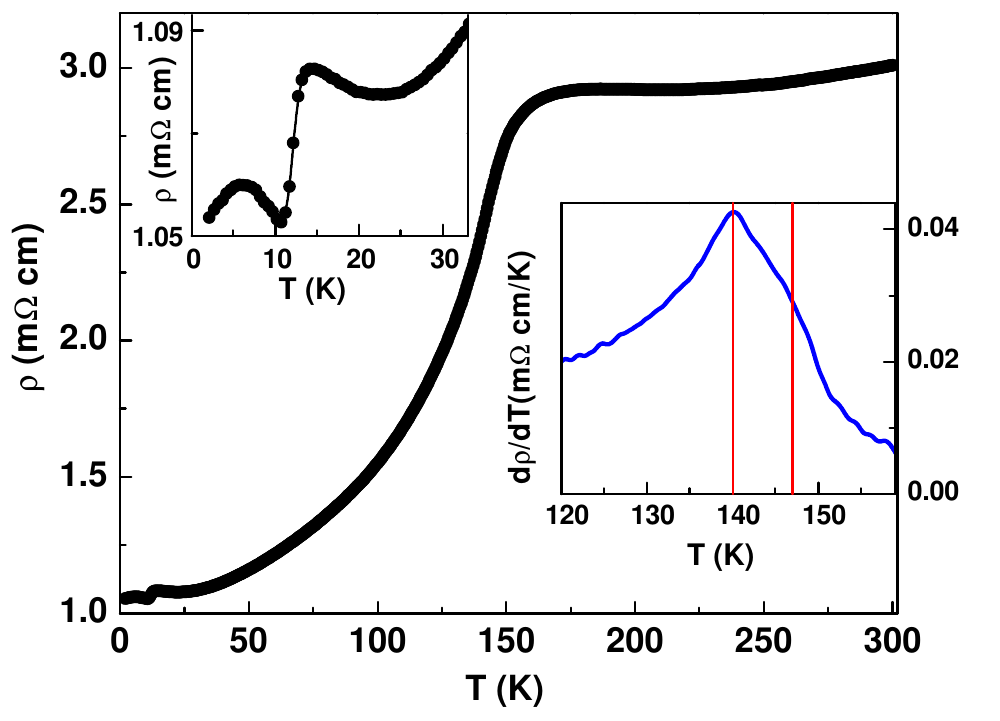}\\
\caption{(Color online) Temperature dependence of resistivity for the PrFeAsO sample in zero magnetic field. Lower inset: d$\rho$/d$T$ versus $T$ curve of
the sample displaying the structural and magnetic transitions at 147 K and 140 K, respectively. Upper inset: The low-temperature behavior of $\rho$ displaying
the resistivity minimum at $T_{min}\sim$ 23 K, AFM ordering of the Pr moments at $T_N^{\rm{Pr}}\sim$ 12 K and the maximum in $\rho$ due to the Fe spin
reorientation at $T_{SR}\sim$ 6 K.}\label{Fig.1}\label{Fig.1}
\end{figure}

After the sharp drop at $T_{N}^{\rm{Pr}}$, $\rho$ starts to increase slowly with the decrease of temperature and then passes through a maximum at $T_{SR}$=6 K.
This behavior is quite unusual in metallic antiferromagnets. Normally, the resistivity below AFM ordering continues to decrease due to the reduction of magnetic
scattering with decreasing temperature. In this context, it is important to mention that $\rho$($T$) of other $R$FeAsO pnictides with magnetic rare-earth ion does not exhibit any such anomaly below the AFM ordering of $R$ moments \cite{tian,jesche,cheng,mart}. This unique feature of PrFeAsO compound shows that the Pr
ordering has an influence on the charge scattering mechanism which is quite different in nature from that observed in other pnictides with magnetic $R$ ion. In
$R$FeAsO, the detailed comparison of the experimental data from $\mu$SR, neutron and M\"{o}ssbauer spectroscopy shows that the strength and nature of magnetic
coupling between iron and magnetic rare-earth sublattices are sensitive to $R$ ion \cite{maeter}. The coupling is strongest in the case of Ce where the Ce-Fe
exchange interaction is of the same order as the Ce-Ce interaction while the $R$-Fe interaction is relatively weak for $R$=Sm and Pr \cite{maeter}. Hence, only
the strength of the $R$-Fe interaction is not enough to explain the observed resistivity anomaly at 6 K in PrFeAsO. From the M\"{o}ssbauer spectroscopy data of
$R$FeAsO, McGuire {\it et al}. \cite{mcguire1} determined the nature of spin ordering of the $R$ ion and its influence on the in-plane Fe moments ordering. They
have shown that below $T_{N}^{R}$, the Ce moments in CeFeAsO order nearly in the $ab$ plane while the Pr moments in PrFeAsO lie along the $c$ axis. This specific ordering of Pr sublattice causes a Fe spin reorientation below $T_{N}^{\rm{Pr}}$ with a significant component of the Fe spin along the $c$ axis. No such
reorientation of the Fe spins occurs below $T_{N}^{R}$ in the Sm, Nd and Ce compounds \cite{maeter,mcguire1}. Based on the electronic potential calculations
and symmetry analysis, a qualitatively similar model was proposed by Maeter {\it et al} \cite{maeter}. This spin-reorientation model explains the increase in
quadrupole shift in M\"{o}ssbauer spectroscopy \cite{mcguire1} and the decrease in $\mu$SR frequency \cite{maeter} below $T_{N}^{\rm{Pr}}$ for PrFeAsO.

We can use the same model to explain the observed resistivity increase below $T_{N}^{\rm{Pr}}$. As the Fe layer is responsible for charge conduction mechanism,
any spin reorientation within the Fe sublattice will affect the charge scattering process, which in turn change the resistivity. Since the spin reorientation
phenomenon evolves slowly as a second order phase transition immediate after the AFM ordering of Pr moments, one expects an increase in resistivity just below
$T_{N}^{\rm{Pr}}$ due to this additional scattering. However, at temperatures below $T_{SR}$ when this ordering is complete, resistivity is expected to decrease
with decreasing $T$. Qualitatively, the dependence of $\rho$ on $T$ in the vicinity of $T_{SR}$ would be similar to those observed around $T_N$ and
$T_{N}^{\rm{Pr}}$. It may be mentioned that earlier resistivity data on PrFeAsO do not show anomaly at $T_{SR}$ \cite{mcguire1,rotundu}. Though
McGuire {\it et al}.\cite{mcguire1} observed the Fe-spin reorientation from the M\"{o}ssbauer spectroscopy, the $\rho$($T$) curve of their sample does not
show any anomaly below $T_N$. We believe that the manifestation of the Fe-spin reorientation phenomenon is particularly weak in the resistivity and sensitive
to sample quality.

\subsection{Magnetoresistance}
\begin{figure}[hb]
\includegraphics[width=0.42\textwidth]{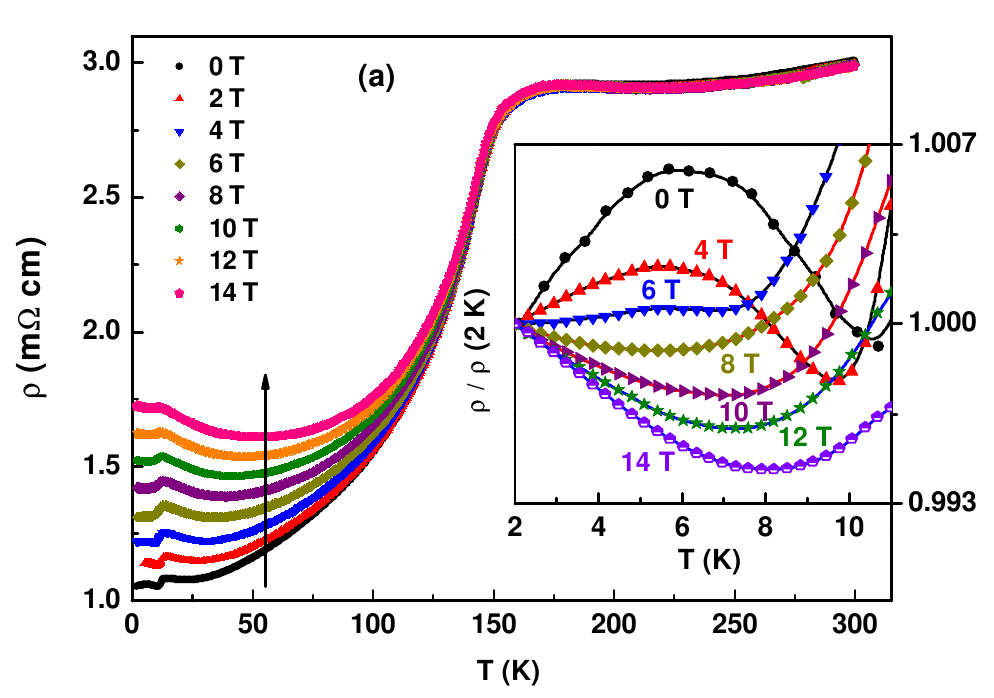}\\
\includegraphics[width=0.41\textwidth]{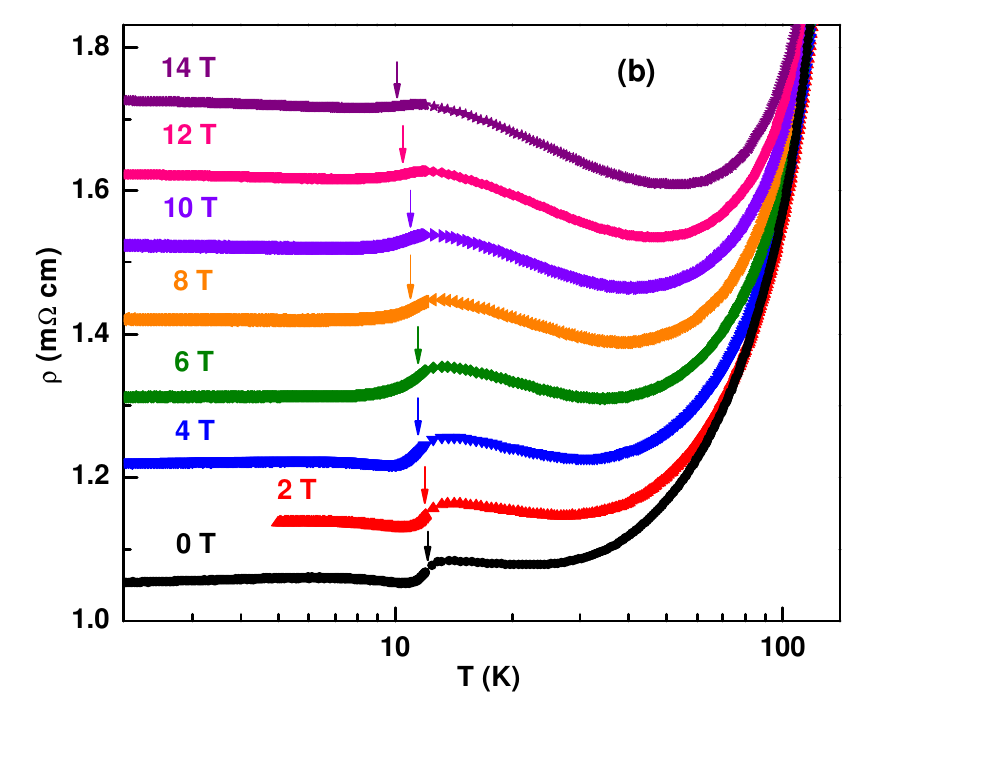}\\
\includegraphics[width=0.4\textwidth]{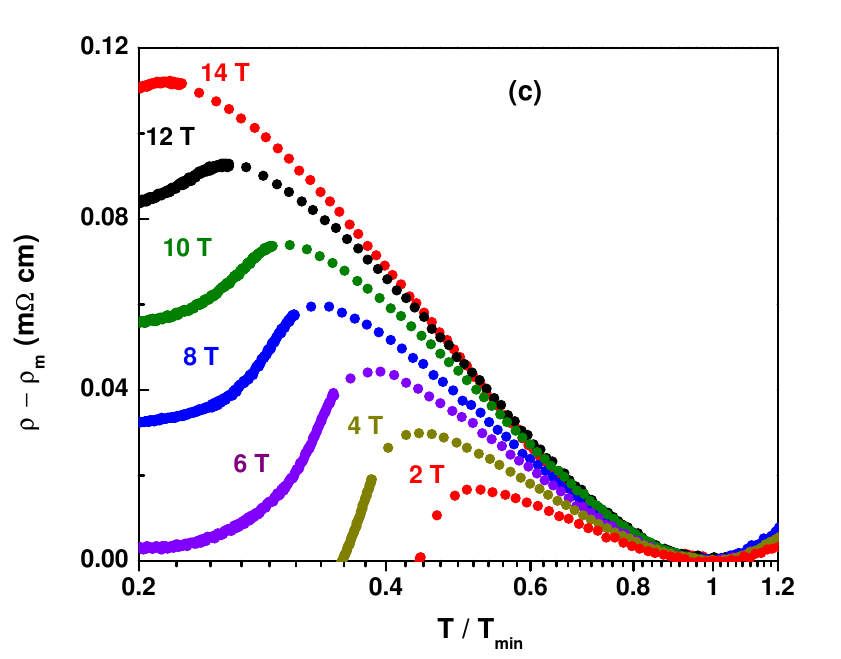}\\
\caption{(Color online) (a) Temperature dependence of the resistivity for the PrFeAsO sample at different magnetic fields. The arrow indicates the direction of
increase of the magnetic field. Inset shows the effect of applied magnetic field on the maximum in $\rho$ at $T_{SR}$. The resistivity has been normalized to
its value at 2 K for clarity. (b) Semi-logarithmic plot of the $\rho$ versus $T$ curve below 140 K for different magnetic fields. Arrows mark the $T_N^{Pr}$ in
different magnetic field. (c) The normalized log$T$ behavior of resistivity wherein ($\rho-\rho_m$) has been plotted against $T/T_{min}$.}\label{Fig.2}
\end{figure}
In order to investigate the effect of magnetic field ($H$) on the AFM Pr and Fe sublattices, we have measured the temperature dependence of resistivity of the
PrFeAsO sample for different $H$ [Fig. 2(a)]. The application of magnetic field up to 14 T does not affect the structural and magnetic transitions around 150 K.
On the other hand, both resistivity and magnetic transitions well below 150 K are strongly influenced by the magnetic field. The transition at $T_{N}^{\rm{Pr}}$
gets smeared with the increase in $H$; $T_{N}^{\rm{Pr}}$ decreases by 1.9 K and the transition width increases from 2 to 3.6 K as $H$ increases from 0 to 14 T.
The maximum at $T_{SR}$ is also sensitive to $H$ as shown in the inset of Fig. 2(a). It broadens and shifts towards the lower temperature with the increase of
$H$ up to 6 T. Above 6 T, this maximum in resistivity disappears but the upturn remains.

The weak upturn observed in resistivity below $T_{min}$ at zero-field is significantly enhanced with applied magnetic field and $\rho$ increases approximately as logarithm of temperature in the region $T_{N}^{\rm{Pr}}$$<$$T$$<$$T_{min}$. Figure 2(b) shows the temperature dependence of $\rho$ in semi-logarithmic scale below 140 K for different $H$. It is clear from the figure that $T_{min}$ shifts towards the higher temperature side from 23 to 55 K and the value of the resistivity ($\rho_{m}$) at the minimum increases from 1.07 to 1.6 m$\Omega$ cm as the magnetic field increases from 0 to 14 T; both $T_{min}$ and $\rho_{m}$ show an approximate linear dependence on $H$. Figure 2(c) normalizes the log$T$ behavior of resistivity wherein ($\rho-\rho_{m}$) is plotted against $T/T_{min}$. It may be mentioned that $\rho$($T$) below $T_{N}^{\rm{Pr}}$ also shows log$T$ dependence for $H$$\geq$12 T. Normally, the log$T$ dependence of resistivity
is consistent with the Kondo scattering. In Kondo effect, magnetic field suppresses the conventional spin-flip scattering and the minimum shifts towards lower
temperature side. However, in the present case $T_{min}$ shifts towards higher temperature side and the upturn in resistivity below $T_{min}$ is enhanced with the increase of $H$ which go against the Kondo scenario.

As the field dependence of magnetoresistance (MR) may provide important information about the origin of large value of MR, we have measured $\rho$ as a function of $H$ at different temperatures which were reported in our earlier publication [Fig. 3]\cite{bhoi2}. Above 150 K, the value of MR is very small ($\leq$1\%) even at a field of 14 T. Below 150 K, the MR increases rapidly with decreasing temperature and it reaches 63\% at 5 K and 14 T as shown in the inset of Fig. 3. It is
interesting to note that MR for PrFeAsO at 5 K increases linearly with $H$ up to 6.4 T above which MR also increases linearly but with a higher slope, signalling a weak metamagnetic transition. Earlier we have shown that the maximum at 6 K in the $\rho$($T$) curve also disappears above 6 T. MR increases linearly with $H$ in the entire field range below 40 K. For $T \leq$ 40 K, MR versus $H$ curve develops a weak curvature in the low-field region which indicates a crossover from $H$ linear to $H^2$ dependence as $H\rightarrow$0. The $H$ linear MR originates from the Dirac cone states and has been explained by the quantum mechanical model proposed by Abrikosov \cite{bhoi2,rich,abri}.

\begin{figure}[h]
\includegraphics[width=0.45\textwidth]{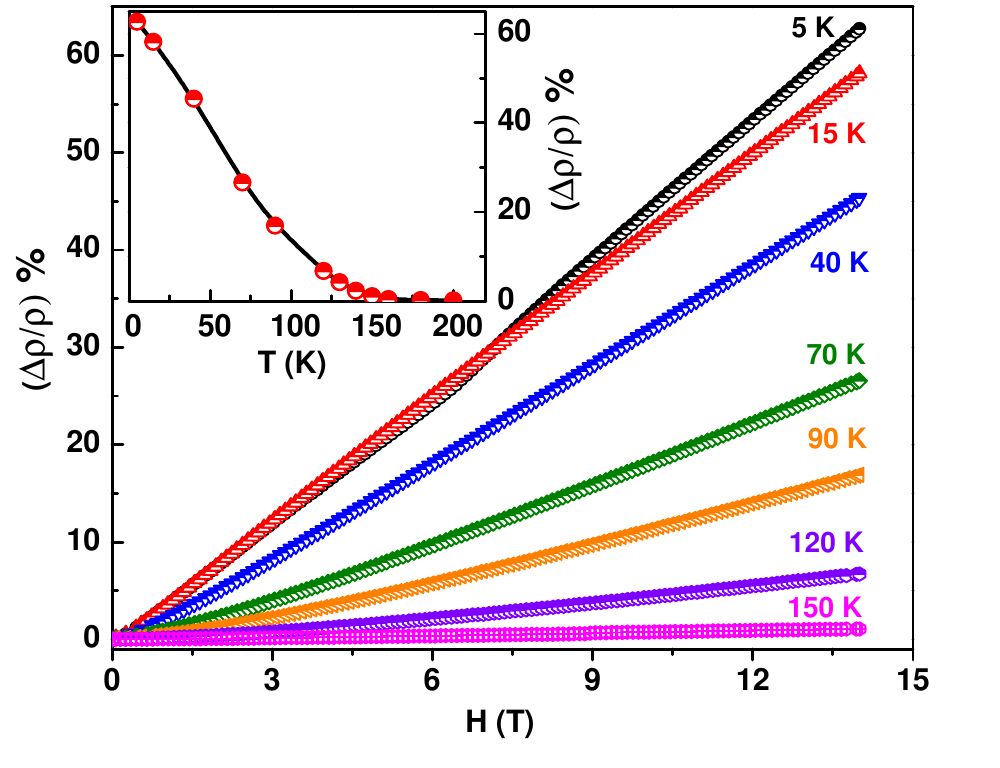}\\
\caption{(Color online) Field dependence of the MR [=$\Delta\rho/\rho(0)$] for the PrFeAsO sample at different temperatures. Inset: Temperature dependence of
the MR at 14 T.}\label{Fig.3}
\end{figure}

The observed MR in PrFeAsO is comparable to that of NdFeAsO and CeFeAsO compounds having 4$f$ electrons \cite{cheng,jesche}. However, the value of MR in LaFeAsO
is 30$\%$ at 5 K and 14 T \cite{dong}, which is half of the MR observed in our case. This suggests that the 4$f$ electrons of rare-earth ion also contribute in
the electronic scattering process. The application of external magnetic field enhances the magnetic moment fluctuations in one of the two AFM sublattices of Pr
which is antiparallel to $H$. This in turn increases the spin-disorder scattering. The reduction of staggered magnetization of AFM Pr sublattice with applied
field is the possible reason for the disappearance of the maximum at $T_{SR}$. Since this peak originates due to the induction by AFM Pr sublattice, the
disordering in the AFM Pr sublattice may reduce the out-of-plane Pr-Fe exchange interaction considerably which is responsible for Fe spin reorientation.
As the electronic and magnetic properties of pnictides are anisotropic in nature, the observed resistivity is an average effect over all grains with different
orientations. When a magnetic field is applied, the response seen is some average on all grains in the sample that depends on the angle between the applied
magnetic field and the easy axis of magnetization. This averaging might be the one of the reasons for broadening the peak with field. In order to avoid this,
studies of magnetotransport properties along different crystallographic directions on high quality single crystal will be useful. However, due to the
nonavailability of PrFeAsO single crystal of appropriate size, transport and magnetic studies on single crystal are not yet reported. In such a situation,
we believe, the present results are important for understanding the role of 4$f$ Pr moments on charge carriers.
\begin{figure}[h]
\includegraphics[width=0.45\textwidth]{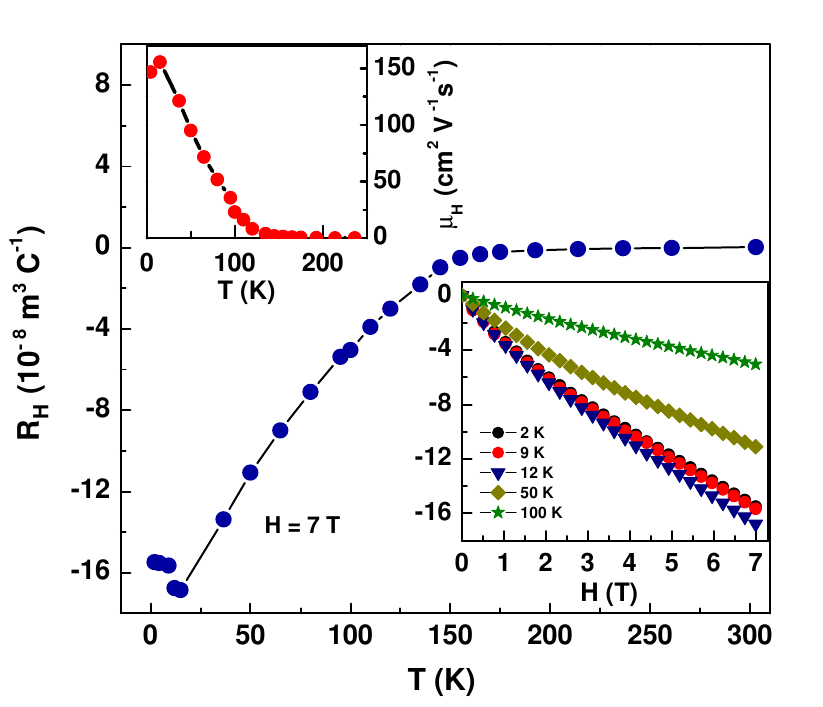}\\
\caption{(Color online) Temperature dependence of the Hall coefficient $R_H$ at $H$ = 7 T and Hall mobility $\mu_H$ (upper inset) for the PrFeAsO sample.
The lower inset shows magnetic field dependence of $R_H$ for some selected temperatures.}
\label{Fig.4}
\end{figure}
For further understanding the role of Pr 4$f$ electron on charge scattering, we have measured the transverse magnetoresistance of the sample at different
temperatures. Figure 4 shows the temperature dependence of the Hall coefficient $R_H$. The negative value of $R_H$ confirms that electron plays dominant role
in the conduction process. We observe that at high temperature $R_H$ is linear in $H$ but it shows a weak nonlinear $H$ dependence at low temperature
(lower inset of Fig. 4) as in the case of NdFeAsO single crystal \cite{cheng}. Similar to other $R$FeAsO compounds \cite{mcguire1,cheng}, $R_H$ is almost
temperature independent in the paramagnetic state above 150 K but increases rapidly in magnitude upon cooling in the SDW AFM state. Like resistivity, the
temperature dependence of Hall coefficient shows a pronounced anomaly around the AFM ordering of the Pr moments and a change in slope below $T_{SR}$. This
evidences an interaction between Pr 4$f$ and Fe 3$d$ electrons in the present system. This is also clear from the sudden drop in the Hall mobility
$\mu_H$ ($=R_{H}\rho^{-1}$) below $T_{N}^{\rm{Pr}}$ (upper inset of Fig. 4). Normally, $R_H$ changes significantly at the order-disorder magnetic phase
transition when the same electron takes part both in magnetism and electrical transport. However, in the present case Fe layer is responsible for the charge
conduction and so the manifestation of the effect of $R$ ordering on the Hall coefficient is rather unusual and suggests the existence of intersublattice
magnetic coupling.

\subsection{Specific heat}

The specific heat for $R$FeAsO compounds has been reported for various $R$ elements wherein the manifestations of magnetic ordering of Fe and $R$ moments have been noticed \cite{ding,mcguire1,jesche,dong,tian,riggs1}. However, the specific heat for PrFeAsO  down to or below $T_{N}^{\rm{Pr}}$ has not been reported yet. We have measured the specific heat for PrFeAsO compound down to 2 K in order to investigate whether the Fe spin reorientation  has any effect on the same. Figure 5(a) shows the temperature dependence of  $C$ from 200 to 2 K for different $H$. In the temperature range 135-155 K, the $C(T)$ curve exhibits a broad anomaly, which is related to the structural transition and the AFM ordering of the Fe moments. Similar to $\rho(T)$, the specific heat at these transitions are  insensitive to $H$. Two overlapping peaks at 140 and $\sim$ 147 K are clearly visible in the background-subtracted specific heat [upper inset of Fig. 5(a)] similar to that reported earlier \cite{mcguire1} and in good agreement with the temperatures at which the peaks in the d$\rho$/d$T$ curve occur. The total entropy change associated with both the transitions is calculated to be 0.4 J mol$^{-1}$K$^{-1}$ by integrating the background-subtracted heat capacity data. Another important feature in the $C(T)$ curve is a sharp anomaly at 12.1 K due to the AFM ordering  of the Pr moments [lower inset Fig. 5(a)]. Unlike resistivity, no visible anomaly is reflected in the $C$($T$) curve below $T_N^{\rm{Pr}}$ due to the spin reorientation of the Fe moments. The reasons may be that the spin reorientation phenomenon evolves as a weak second-order phase transition and it begins with the onset of the relatively stronger rare-earth magnetic ordering. Due to the proximity of these two orderings, the effect of the weaker ordering on specific heat may be masked by the stronger one. That is why the effect of Fe spin reorientation has not been detected earlier in $\mu$SR, neutron scattering and magnetic susceptibility studies \cite{maeter}. In such cases, the derivative curve may disclose any possible effect of the Fe spin reorientation on the specific heat curve. In the derivative of $C/T$ (or $C$) data at $H$ = 0  [Fig. 5(b)], we observe a broad peak at 7 K which is well separated from the strong anomaly at 12.1 K. Thus,  the spin reorientation do have some influence on the specific heat. Similar to resistivity, on application of the magnetic field, the peak becomes broader. However, there is one important difference between the field dependence of $\rho$ and $C$. The peak in $\rho$($T$) at $T_{SR}$ is very sensitive to the applied magnetic field and it is no more visible above a moderate field strength $\sim$ 6 T. This is not the case for specific heat where the peak is very much noticeable up to 14 T.
\begin{figure}[h]
\includegraphics[width=0.5\textwidth]{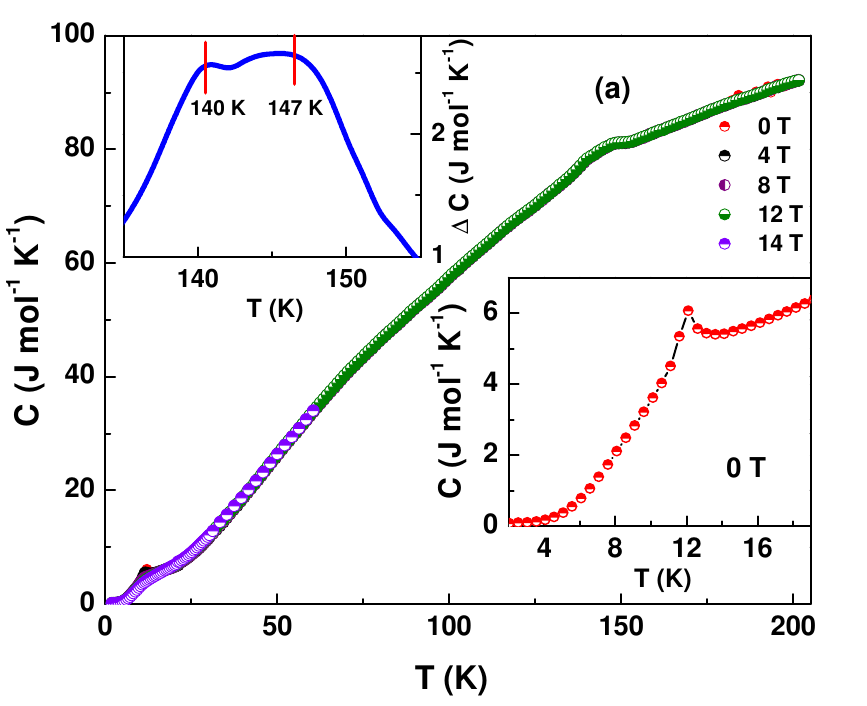}\\
\includegraphics[width=0.5\textwidth]{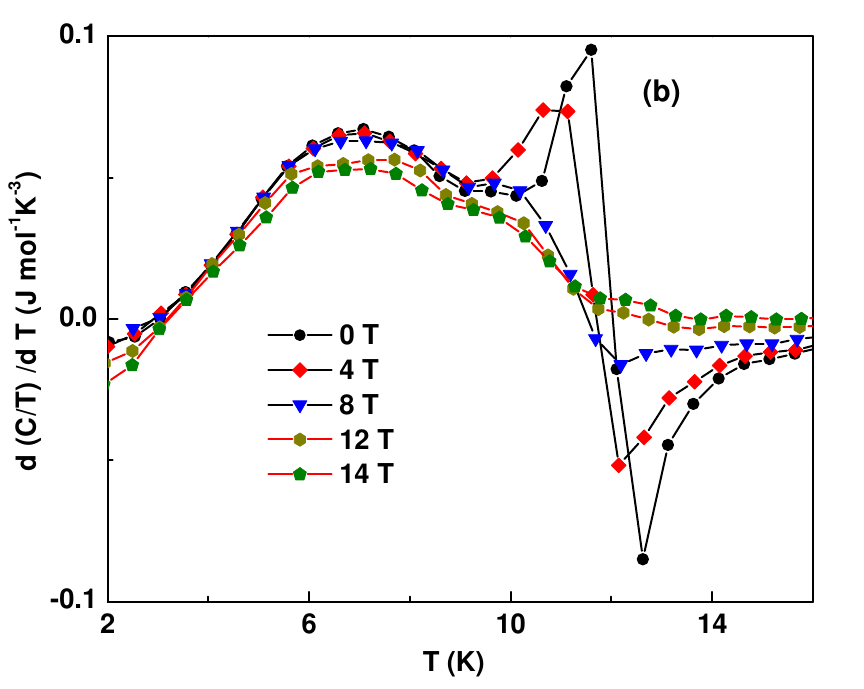}\\
\caption{(Color online) (a) Temperature dependence of the specific heat $C$ for the PrFeAsO sample at different fields. Upper inset: The cusp-like feature in
the background subtracted specific heat data at the structural (147 K) and magnetic (140 K)  transitions. Lower inset: The peak at the AFM ordering of the Pr
moments. (b) Temperature dependence of the derivative of $C/T$ curve for different magnetic fields.}
\label{Fig.5}
\end{figure}
In order to evaluate the lattice and electronic parts in the heat capacity, the data are fitted to the equation:
\begin{equation}
C_b(T) = \gamma T + A_{D}C_{D}(T,\theta_D) + A_{E}C_{E}(T, \theta_E)
\end{equation}
where $\gamma$ is the Sommerfeld coefficient, and C$_D$ and C$_E$ are the Debye and Einstein functions, respectively. Fits were carried out over the temperature
range 50 K $\leq T \leq$ 200 K  excluding the transition region 120 K $\leq T \leq$ 160 K where the anomaly in the specific heat occurs. The parameters derived
from the above fitting are $\gamma \sim$ 34 mJ mol$^{-1}$ K$^{-2}$, $A_D \sim $ 152 J mol$^{-1}$ K$^{-2}$, $A_E \sim$ 55 J mol$^{-1}$ K$^{-2}$, $\theta_D \sim$
212 K, and $\theta_E\sim$ 445 K. The calculated value of $\gamma$ for the present sample is comparable to those observed in SmFeAsO
($\gamma^{\rm{Sm}}$ = 42 mJ mol$^{-1}$ K$^{-2}$) and CeFeAsO ($\gamma^{\rm{Ce}}$ = 36 mJ mol$^{-1}$ K$^{-2}$) compounds \cite{riggs1}. However, these values of
$\gamma$  are one order of magnitude larger than that for LaFeAsO ($\gamma^{\rm{La}}$ = 3.7 mJ mol$^{-1}$ K$^{-2}$) \cite{dong}. The relatively high value of
$\gamma$, sharp drop in resistivity and the anomaly in the Hall coefficient at $T_N^{\rm{Pr}}$ confirm some kind of interaction between the conduction electrons in Fe layer and the localized 4$f$ electrons of the Pr ion. Normally, the heat capacity of an antiferromagnet contains a magnetic component due to the SDW excitations. The typical SDW contribution to $C$ for the AFM material $C_{SDW}$ $\propto$ $T^3$, is assumed to be small and cannot be extracted from our data. This is also the case for other undoped pnictides where the background is calculated using lattice and electronic contributions only \cite{dong,riggs1}. We have also measured the heat capacity for the superconducting PrFeAsO$_{1-x}$F$_x$ sample to calculate the background and observed that both the methods reveal almost same result.

The background-subtracted specific heat data for the undoped PrFeAsO is plotted in Fig. 6(a) as a function of temperature for 2 K $\leq T \leq$ 40 K for different applied magnetic fields. At $H$=0, the anomaly is quite sharp with a jump $\Delta C$$\sim$5 J mol$^{-1}$ K$^{-1}$. The entropy removal related to this magnetic transition can be evaluated as $S_m = \int(C_m/T)dT$, where $C_m = C - C_b$ and $C_b$ is the sum of the electronic and lattice parts as described in Eq.(2). For PrFeAsO, $S_m$ tends to saturate at $\sim$4.2 J mol$^{-1}$ K$^{-1}$ [inset of Fig. 6(a)], which is 70$\%$ of the entropy associated with the AFM spin ordering for a doublet ground state (R$\ln$2). The large value of the entropy change leads us to believe that the low-temperature peak is essentially due to the long-range AFM ordering of the Pr moments. Unlike $T_S$ and $T_N$, $T_N^{\rm{Pr}}$ is  sensitive to the external magnetic field.  The peak at $T_N^{\rm{Pr}}$  broadens with the  increase of $H$. The role of $H$ on  this magnetic transition may also be understood from the entropy change at $T_N^{\rm{Pr}}$. Though, the calculated values of $S_m$ for different magnetic fields are close to the zero-field one, the entropy change at $T_N^{\rm{Pr}}$ decreases with the increase of $H$. This indicates that the magnetic field induces disorder in the AFM Pr sublattice. The nature of $H$ dependence of $C$ at $T_N^{\rm{Pr}}$ is similar to that observed in CeFeAsO but different from SmFeAsO. In  CeFeAsO, the peak width increases appreciably  with the increase of $H$, whereas in SmFeAsO the peak remains quite sharp even at a field as high as 16 T \cite{riggs1}.
\begin{figure}[t]
\includegraphics[width=0.5\textwidth]{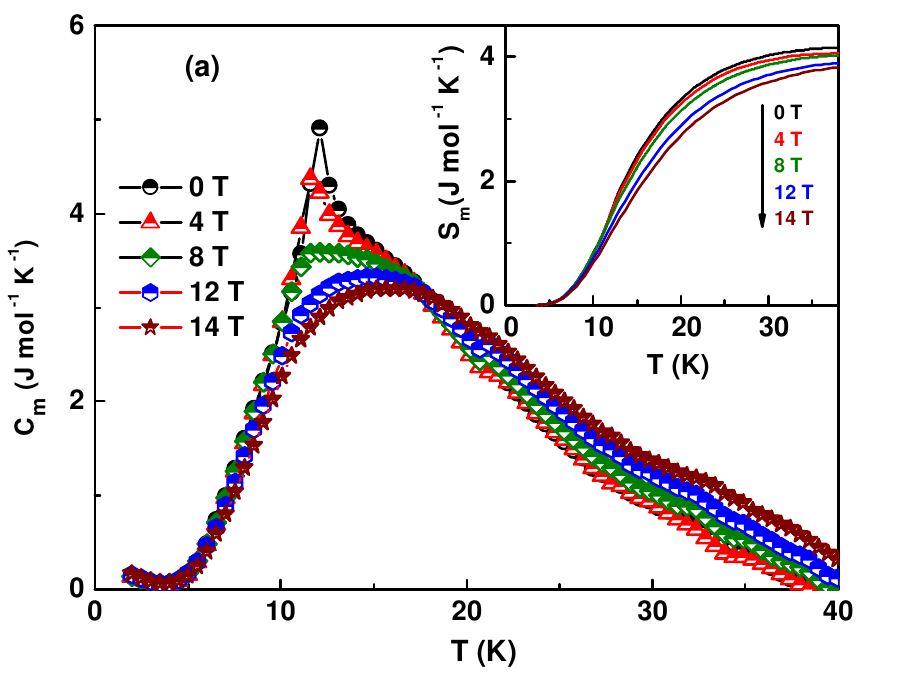}\\
\includegraphics[width=0.5\textwidth]{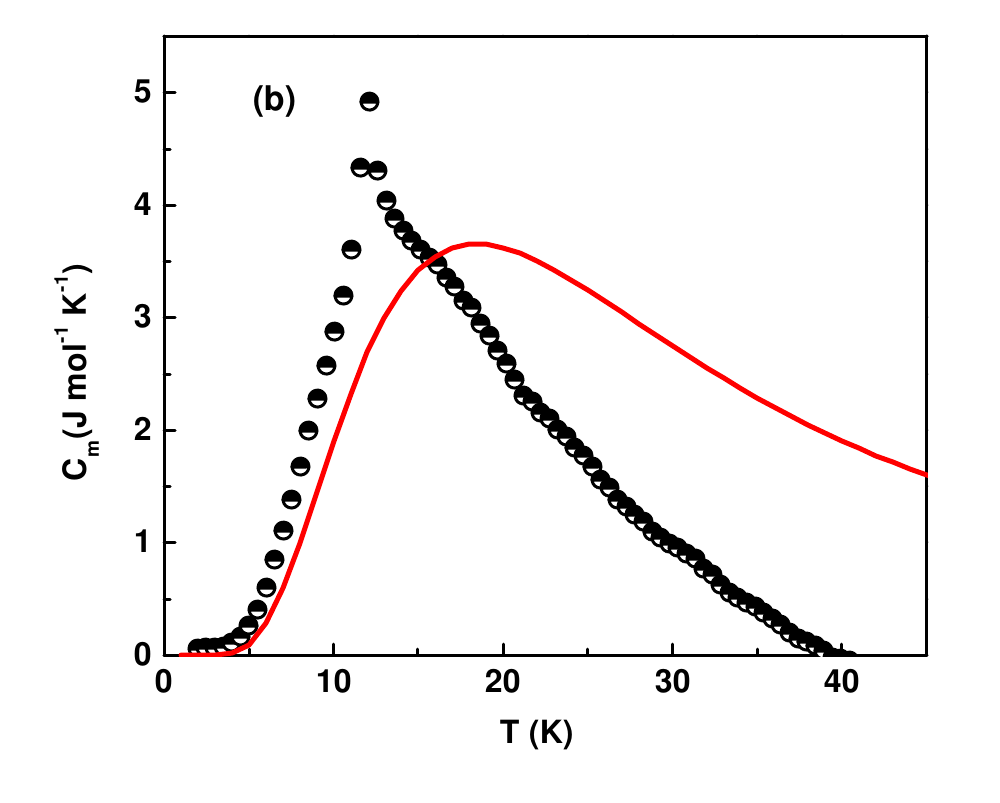}\\
\caption{(Color online) (a) Background subtracted specific heat data in the temperature range 2-40 K for different magnetic fields. Inset: Entropy removal
$S_m$ related to the magnetic transition for different magnetic fields. (b) The calculated specific heat versus $T$ curve using the Scottky energy levels proposed for PrFePO in Ref.\cite{baumb} is plotted along with the experimental data for PrFeAsO.}
\label{Fig.6}
\end{figure}
The observation of the pronounced peak at 12.1 K seems in contradiction with that of the Schottky type anomaly observed in the specific heat of PrFePO single
crystals with tetragonal $P4/nmm$ symmetry \cite{baumb}. For PrFePO, the broad anomaly in specific heat data was  described by a two-level scheme arising from
the energy difference between the crystalline electric field (CEF) ground state and the first excited state. Analysis of the specific heat and susceptibility data suggests that the energy separation between the ground and first excited states in PrFePO is $\sim$ 44.3 K and both the states are nonmagnetic singlet \cite{baumb}. We have calculated specific heat using the two-level Schottky model as proposed for PrFePO \cite{baumb} and plotted in Fig. 6(b) along with the  experimental data for the PrFeAsO sample from 2 to 40 K. The calculated curve shows the expected broad Schottky-type anomaly with maximum value of $C$,
$C_{max}$ = 3.64 J mol$^{-1}$ K$^{-1}$ at $T$ = 18.4 K. This is in contrast to the experimental result where a sharp anomaly with
$C_{max}$ = 5 J mol$^{-1}$ K$^{-1}$ at 12.1 K is observed.  In this context, we would like to mention that the low-temperature anomaly in the specific heat of
Ce- and Sm-based compounds has also been attributed to the long-range AFM ordering of rare-earth moments \cite{jesche,torp,riggs1}.

Both the transport and specific heat data of PrFeAsO exhibit multiple structural and magnetic phase transitions. The structural transition at $T_S$ and the AFM
ordering of Fe at $T_N$ do not change with the application of magnetic field, while the AFM ordering of the Pr ion at $T_{N}^{\rm{Pr}}$ and the spin reorientation transition at $T_{SR}$ are significantly affected by magnetic field. The split lattice-magnetic transitions 140-147 K suggest a moderate spin-phonon coupling. However, one may consider a noticeable electron-magnon-phonon coupling for the low temperature transitions. Because of a strong electron-magnon coupling, the magnetic field could have a big influence on the electronic system (large magnetoresistance), while a weaker magnon-phonon coupling would not change the lattice system (i.e. $C(T)$) too much. On the other hand, the appearance of a sharp peak at $T_{N}^{\rm{Pr}}$ in the $C(T)$ curve and its rounding off at high magnetic field indicates that the spin-lattice coupling is rather significant. Indeed, neutron diffraction studies in PrFeAsO reveals an anomalous expansion of the $c$ axis below 40 K leading to a negative thermal expansion of the volume which is attributed to the magnetoelastic coupling between the ordered Fe and Pr sublattices \cite{kimber}. To estimate the relative strength of different couplings further studies such as effect of pressure on resistivity, magnetization and Raman spectra might be helpful.

\section{conclusion}

In summary, we have investigated the effect of magnetic field on the structural and magnetic transitions, and the influence of Pr moments ordering on conduction
electrons in PrFeAsO from resistivity, Hall coefficient and specific heat measurements. Anomalies are clearly reflected in $\rho(T)$, $R_H(T)$ and $C$($T$) curves at the structural transition ($T_S$), and at the AFM ordering temperatures of Fe ($T_N$) and Pr ($T_N^{\rm{Pr}}$) moments. The $\rho(T)$ and $C$($T$) data at $T_S$ and $T_N$ are not affected by the magnetic field $H$. However, below $T_N$, $\rho$ is significantly enhanced by the magnetic field. The width of transition at $T_N^{\rm{Pr}}$ in $\rho$($T$) and $C$($T$) increases appreciably with the increase of $H$ due to the reduction of staggered magnetic moment. The sharp anomaly in $C$($T$) at $T_N^{\rm{Pr}}$ and the large value of entropy removal $S_m$ = 4.2 J mol$^{-1}$ K$^{-2}$ (70\% of R$\ln$2) related to this transition indicate that most of the degrees of freedom of lowest CEF doublet are involved in the AFM order.

Both $\rho(T)$ and $d(C/T)/dT$ curves exhibit a broad peak at $T_{SR}$ (6-7 K) as a result of the spin reorientation in Fe sublattice induced by the AFM
ordering of Pr moments. The increase of $\rho$ just below $T_N^{\rm{Pr}}$ suggests that Fe spin reorientation enhances the charge scattering process, an evidence for the influence of Pr moments on conduction electrons in PrFeAsO. Unlike other magnetic rare-earth ions, the moments of Pr ions order along the $c$ axis. This
specific ordering of Pr moments affects the ordered in-plane Fe moments significantly below $T_N^{\rm{Pr}}$ and induces a  component of the Fe spin along the
$c$ axis. Such reorientation of the Fe spins does not occur in the Sm, Nd and Ce compounds. Even in PrFeAsO, as this phenomenon is weak and occurs in the proximity of the relatively stronger Pr-ordering transition, its effect has not been detected earlier from the neutron diffraction, magnetic susceptibility and other measurements.

Our results on magnetoresitance and the Hall coefficient also give important information related to the charge conduction mechanism and the interaction between
Pr moments and conduction electrons. $R_H$ is large, negative and shows strong temperature dependence in the SDW AFM state. The negative sign of $R_H$ indicates
that the electron plays an important role in charge conduction process. The large value of MR in $R$FeAsO with magnetic $R$ ions and the strong anomaly in $R_H$
and $\mu_H$ at $T_N^{\rm{Pr}}$ are further evidences for the coupling between  rare earth 4$f$ electrons and conduction electrons.

Three couplings: spin-lattice, spin-electron and electron-phonon of different magnitudes dominate the physical properties of PrFeAsO. The spin-lattice coupling
appears to be weak at high temperature, while at low temperature the strength of spin-lattice and electron-magnon coupling is quite significant.

\section{Acknowledgement}

The authors would like to thank A. Pal and S. Banerjee for technical help during sample preparation and measurements. We would also like to thanks A. Midya and
N. Khan for their help during the Hall effect measurements. V. Ganesan would like to thank DST, India for financial assistance for the 14 T physical property measurement system facility at UGC-DAE CSR, Indore.

\end{document}